\documentclass[5p,times,sort&compress]{elsarticle}

\usepackage{url}

\usepackage{amsmath}
\usepackage[utf8]{inputenc}
\usepackage[breaklinks]{hyperref}

\usepackage{graphicx}

\hypersetup{colorlinks=true, linkcolor=blue!50!black, urlcolor=blue!50!black, citecolor=blue!50!black}
\journal{Computer Physics Communications}

\begin{document}

\author{Pierre de Buyl\corref{kul}}
\cortext[kul]{Present address: Department of Chemistry, Katholieke Universiteit Leuven, Celestijnenlaan 200F, B-3001 Heverlee, Belgium}
\address{Center for Nonlinear Phenomena and Complex Systems, Université libre de Bruxelles, CP231, Campus Plaine, 1050 Brussels, Belgium}
\title{The vmf90 program for the numerical resolution of the Vlasov equation for
mean-field systems}
\date{\today}

\begin{abstract}
  The numerical resolution of the Vlasov equation provides complementary
  information with respect to analytical studies and forms an important
  research tool in domains such as plasma physics. The study of mean-field
  models for systems with long-range interactions is another field in which the
  Vlasov equation plays an important role. We present the vmf90 program that
  performs numerical simulations of the Vlasov equation for this class of mean-field
  models with the semi-Lagrangian method.
\end{abstract}


\maketitle


{\bf PROGRAM SUMMARY}

\begin{small}
\noindent
{\em Manuscript Title:} The vmf90 program for the numerical resolution of the
Vlasov equation for mean-field systems \\
{\em Author:} Pierre de Buyl \\
{\em Program Title:} vmf90 \\
{\em Journal Reference:}                                      \\
{\em Catalogue identifier:}                                   \\
{\em Licensing provisions:} vmf90 is available under the GNU General Public License \\
{\em Programming language:} Fortran 95 \\
{\em Computer:} Single CPU computer \\
{\em Operating system:} No specific operating system, the program is tested
under Linux and OS X \\
{\em RAM:} about 5 M bytes 
{\em Classification:} 1.5 Relativity and Gravitation, 19.8 Kinetic Models,
19.13 Wave-plasma Interactions, 23 Statistical Physics and Thermodynamics. \\
{\em External routines/libraries:} HDF5 for the code (tested with HDF5 v1.8.8
and above). Python, NumPy, h5py and matplotlib for analysis. \\
{\em Nature of problem:}\\
Numerical resolution of the Vlasov equation for mean-field models (Hamiltonian
Mean-Field model and Single Wave model).\\
{\em Solution method:}\\
The equation is solved with the semi-Lagrangian method and cubic spline
interpolation. \\
{\em Running time:}\\
The examples provided with the program take 1m30 for the Hamiltonian-Mean Field
model and 10m for the Single Wave model, on an Intel Core i7 CPU @
3.33GHz. Increasing the number of grid points or the number of time steps
increases the running time. \\
\end{small}

\section{Introduction}

The Vlasov equation provides an appropriate description of collisionless
plasmas, beam propagation, collisionless gravitational dynamics, among other
systems of interest. Another field of application is the study of mean-field
models for long-range interactions~%
\cite{chavanis_hmf_epjb_2006,chavanis_II,campa_et_al_phys_rep_2009,bouchet_et_al_physica_a_2010}
In these systems, thermodynamical and
dynamical properties can differ strongly with respect to short-range interacting
systems.  The understanding of the kinetic properties in these systems relies
importantly on their Vlasov description. Theoretical considerations and $N$-body
simulations form the majority of the literature but numerical simulations of the
Vlasov equation are found in a small number of references~%
\cite{antoniazzi_califano_prl,de_buyl_et_al_prstab_2009,de_buyl_cnsns_2010,%
de_buyl_et_al_self-consistent_pre_2011,de_buyl_et_al_rsta_2011,%
de_buyl_gaspard_pre_2011,de_buyl_et_al_cejp_2012,de_buyl_et_al_thermalization_pre_2013,%
ogawa_yamaguchi_pre_2012}. These simulations provide a
complementary point of view to $N$-body simulations by removing the dependence
on the number of simulated particles. The number $N$ of particles present in a
simulation has a significant impact on the dynamics of these long-range systems,
especially on the timescale of the relaxation to equilibrium. Many studies on
long-range interacting systems are performed on one-dimensional models. This
makes the numerical resolution of the Vlasov equation by Eulerian methods,
instead of particle-in-cell methods, computationally accessible.

Numerical resolutions of the Vlasov equation are being performed since decades,
following the work of Cheng and Knorr~\cite{cheng_knorr_1976} on the time discretization of the
problem. Most of the research in this field is related to the simulation of the
Vlasov-Poisson system in plasma physics. As a consequence, no code is available
for the simulation of simpler mean-field models. Altough existing codes for the
Vlasov-Poisson system could be easily converted for use with mean-field systems,
we propose here a code that is aimed specifically at these systems.

The vmf90 program presented in this article has already been used to perform the
numerical resolution of the Vlasov equation for the Hamiltonian Mean-Field or
free electron laser models in the references~\cite{de_buyl_et_al_prstab_2009}
(at that time, a previous C++ version was used) and \cite{de_buyl_cnsns_2010,%
  de_buyl_et_al_self-consistent_pre_2011,de_buyl_et_al_rsta_2011,de_buyl_gaspard_pre_2011,%
  de_buyl_et_al_cejp_2012,de_buyl_et_al_thermalization_pre_2013}.
The results in the present manuscript have been obtained with version 1.0.0 of
the vmf90 program.

\section{Physical models}

\subsection{The Hamiltonian Mean-Field model}

The Hamiltonian Mean-Field (HMF) model, introduced in Ref.~\cite{antoni_ruffo_1995},
consists of particles interacting via a cosine potential in a periodic box. It
is intended as a simplified description of systems of charged or gravitational
particles in which collective behavior, caused by the all-to-all coupling of
the particles, plays an important role.

The HMF model serves as a reference model in the field of systems with
long-range interactions~\cite{campa_et_al_phys_rep_2009}. It exhibits peculiar
dynamical properties among which is the existence of the so-called {\em
  quasi-stationary states} (QSS). These states occur, starting from an
out-of-equilibrium initial condition, after a $O(1)$ timescale but do not
correspond to the equilibrium predicted by statistical mechanics. The lifetime
of the QSS increases algebraically with the number $N$ of interacting particles
in the system.
In the thermodynamic limit, the dynamics remains stuck in the QSS regime.

The Vlasov equation for the HMF model reads
\begin{eqnarray}
  \label{eq:vlasov}
  \frac{\partial f}{\partial t} &+& p \frac{\partial f}{\partial \theta} - \frac{dV[f]}{d\theta} \frac{\partial f}{\partial p} = 0 ~,\cr
    & & \cr
  V[f](\theta) &=& 1 - m_x[f] \cos\theta - m_y[f] \sin\theta ~,\cr
  m_x[f] &=& \int d\theta dp\ f \cos\theta ~, \cr
  m_y[f] &=& \int d\theta dp\ f \sin\theta ~,
\end{eqnarray}
where $f=f(\theta,p ; t)$ is the one-particle distribution function, $V$ is the
potential, depending self-consistently on $f$ and $m_x$ and $m_y$ are the two
components of the magnetization vector.

The time evolution of Eqs.~(\ref{eq:vlasov}) conserves the following quantities:
the $L_1$ norm
\begin{equation}
  \label{eq:HMF-f1}
  ||f||_1 = \int d\theta\ dp\ f(\theta,p ; t)~,
\end{equation}
the total momentum
\begin{equation}
  \label{eq:HMF-P}
  P(t)[f] = \int d\theta\ dp\ f(\theta,p ; t) ~  p ~,
\end{equation}
and the energy $U$
\begin{align}
  \label{eq:HMF-U}
  U(t)[f] = &\int d\theta\ dp\ f(\theta,p ; t) \cr
  &\left( \frac{p^2}{2} + \frac{1}{2} \left( 1 - m_x[f]\cos\theta - m_y[f]\sin\theta \right) \right) .
\end{align}
These quantities are discussed in Ref.~\cite{de_buyl_cnsns_2010}.

\subsection{The single-wave model}

The so-called single-wave model is a model consisting of many particles
interacting with a single wave. This model originates from the study of
beam-plasma interactions~\cite{OWM_pof_1971} but also proves adequate to study
wave-particle interactions, free electron lasers~\cite{elskens_escande_book}
and collective atomic recoil lasing~\cite{bachelard_et_al_jstat_2010}.

The Vlasov equation for the single wave model reads 
\begin{eqnarray}
  \label{eq:vlasovwave}
  \frac{\partial f}{\partial t} &=& -p \frac{\partial f}{\partial \theta} + 2 (A_x\cos\theta-A_y\sin\theta) \frac{\partial f}{\partial p}~,\cr
  \frac{\partial A_x}{\partial t} &=& -\delta A_y + \int d\theta dp\ f \cos\theta~, \cr
  \frac{\partial A_y}{\partial t} &=& \ \ \delta A_x - \int d\theta dp\ f \sin\theta~,
\end{eqnarray}
where $f=f(\theta,p ; t)$ is the one-particle distribution function and $A_x$
and $A_y$ are the two components of the wave. Specifying $A_x$ and $A_y$ is
equivalent to specifying a phase and an intensity. $\delta$ is a parameter that
accounts for the mismatch between the resonant frequency of the device and the
energy of the particle, it is commonly called the detuning parameter.

The time evolution of Eqs.~(\ref{eq:vlasov}) conserves the following quantities:
the $L_1$ norm
\begin{equation}
  \label{eq:FEL-f1}
  ||f||_1 = \int d\theta\ dp\ f(\theta,p ; t)~,
\end{equation}
the total momentum
\begin{equation}
  \label{eq:FEL-P}
  P(t)[f] = \left(A_x^2+A_y^2\right) + \int d\theta\ dp\ f(\theta,p ; t) ~  p ~,
\end{equation}
where $\left(A_x^2+A_y^2\right)$ is the momentum of the wave, and the energy $U$
\begin{equation}
  \label{eq:FEL-U}
  U(t)[f] = \int d\theta\ dp\ f(\theta,p ; t) \left( \frac{p^2}{2} +
    2 \left( A_y\cos\theta + A_x\sin\theta \right) \right) ~.
\end{equation}

\subsection{Other models}

Other related models can be implemented easily in the code by taking as an
example the HMF model or the single-wave model files.

\section{Description of the vmf90 program}

vmf90 is written in Fortran 90 and takes advantage of user-defined types,
dynamic memory allocation and modules. Fortran 90 provides, with respect to
older revisions of the Fortran programming language, many improvements for the
structured coding of scientific programs. The reader is referred to
Refs.~\cite{decyk_et_al_inheritance_f90_cpc_1998,decyk_gardner_oo_design_cpc_2008,%
  mccormack_fortran_book_2009} for more information on this topic. In addition
to these ideas, vmf90 takes profit of revision control with the git
software~\cite{git-scm.com} and extensive in-code documentation.

\subsection{General structure}

At the core of the program lies the module ``Vlasov\_module'' that provides the
user-defined type (UDT) ``grid'' for phase-space numerical grids. The ``grid''
UDT contains information on:
\begin{enumerate}
\item The number of grid points.
\item The spatial and velocity limits of the grid.
\item Storage for:
  \begin{itemize}
  \item The grid.
  \item The position and velocity marginal distributions.
  \item The second derivative required for the cubic spline interpolation.
  \item The force to apply when evolving the velocities.
  \end{itemize}
\item The periodic or finite character of the spatial dimension.
\item The time step to evolve the system.
\end{enumerate}
The Fortran declaration of the ``grid'' UDT is shown explicitly in
Fig.~\ref{fig:grid}.

This UDT is then used in the description of the HMF and FEL models.
The module ``Vlasov\_module'' also provides the routines to perform the basic
steps for solving the Vlasov equation:
\begin{enumerate}
\item Obtention of the grid coordinates for a given grid point.
\item Advection in the $x$- and $p$-direction.
\item Computation of the distribution marginals in $x$ and $v$.
\item Initialization of the grid with Gaussian or water bag distributions.
\end{enumerate}

For each model, a module provides a UDT containing a ``grid'' variable and
additional information on the model, such as the mean-field variables. A program
using this UDT is written for each model, i.e. ``vmf90\_hmf'' and
``vmf90\_fel'', that sets the simulation up and performs the steps required for
the simulation.
As the code that performs the actual steps in the simulation algorithm is found
in ``Vlasov\_module'', the implementation of other models can benefit from a
verified code base.

A separate module, ``spline\_module'', contains the routines for the cubic
spline interpolation for natural splines (for the velocity dimension and the
non-periodic spatial dimension) and periodic splines (for the spatial dimension
in periodic systems).

The ``vmf90'' module provides information on the code version (see more in
section~\ref{sec:code}).

\begin{figure*}
\begin{verbatim}
  type grid
     integer :: Nx, Nv
     double precision :: xmin, xmax, vmin, vmax
     double precision :: dx, dv
     double precision :: DT
     double precision :: en_int, en_kin, energie, masse, momentum
     logical :: is_periodic

     double precision, allocatable :: f(:,:)
     double precision, allocatable :: d2(:), copy(:)
     double precision, allocatable :: rho(:), phi(:), force(:)
  end type grid
\end{verbatim}
  \caption{The Fortran declaration of the ``grid'' UDT that serves as the
    building block for physical models.\label{fig:grid}}
\end{figure*}

\subsection{Scientific coding guidelines}
\label{sec:code}

With the aim of providing a rigorous scientific computing tool, vmf90 follows
the following principles: use of revision control, i.e. the revision information
is provided in the output file of a vmf90 simulation, no use of global variables
for the computation routines, use of human readable configuration files instead
of hardcoding computation parameters in the program, use of self-describing
data files, code documentation with Doxygen~\cite{doxygen}.
The combination of all these principles allows the scientist to track finely the
code and parameter sets that produce a given result and the scientist-programmer
to maintain a robust codebase for future developments.

The program is under revision control with the git software~\cite{git-scm.com}.
The information on the status of the source code repository is included
automatically during the build process of the software and is updated at each
rebuild so that the executable program always reflect this status.
The programs display revision information: the commit reference (the
SHA1) of the code as well as a status that is one of the following
\begin{itemize}
\item Clean: the code corresponds to the displayed git commit reference.
\item Unclean: the code is based on the displayed git commit reference but has
  been modified.
\item Out of repository: the code has been extracted from a tarball whose
  version is given, but is not tracked by git.
\end{itemize}
In this manner, the results can be linked in a strict manner to the
exact version of the code. It is necessary to use a git-tracked repository to
take profit of this feature.

Due to the use of Fortran UDT's and modules, the full status of a simulated
system is held in a variable. There is thus no need for global variables
(i.e. the common block feature of Fortran is not used, nor are module-global
variables).

All modules are commented in the style of the Doxygen~\cite{doxygen} program. A
full documentation for the API is thus automatically generated.

The program is distributed with the ParseText Fortran module that allows an easy
parsing of human-readable text files for Fortran programs. 
This provides the advantage that the user is less likely to set the wrong
parameters in a configuration file that would contain only numbers without a
close by description. For instance, reading the time step is done with the
following line
of code:
\begin{verbatim}
DT = PTread_d(PT_variable, 'DT')
\end{verbatim}
where {\tt DT} is a double precision variable, {\tt PT\_variable} holds the
information from the configuration file and the argument {\tt 'DT'} indicates
which variable should be looked upon in the configuration file that is
parsed. The routine {\tt PTread\_d} is suited for double precision variables and
similar routines exist for the other common Fortran data types.  A configuration
file is given as an example in the Appendix~\ref{sec:config}.

\subsection{Data storage model and analysis tools}

The vmf90 program stores data in the HDF5 format that allows to store data to
machine precision in structured files that simplify the storage and analysis of
the data. A derivative of the H5MD~\cite{h5md-web,h5md_cpc_2014} file format
is used for that purpose, allowing a consistent storage of parameters and
observables across the file. Also, the version information for vmf90 described
in the previous section is part of the H5MD file.

To allow users who are not familiar with HDF5 or H5MD to analyze the data,
a small program displaying common results for the simulation is provided. It is
a command-line tool that takes as arguments the filename of the data and
a command: ``snaps'' to display snapshots of phase space and ``plot'' to display
observables such as ``mass'' and ``energy''.

This program is written in Python and uses h5py~\cite{h5py2008}, NumPy~\cite{oliphant_cise_2007} and
Matplotlib~\cite{matplotlib}. Its purpose is at the same time to provide a tool to
check on a simulation file as well as an example program that shows how to
access and use the resulting datafile.

\subsection{Obtaining, building and running vmf90}
\label{sec:build}

vmf90 is available under the GNU General Public License, on the GitHub code
hosting website \footnote{\url{https://github.com/pdebuyl/vmf90}}.
In order to have a working version of vmf90 on your computer, here are the steps
to execute in a terminal. Those instructions are provided for the sake of
completeness and to facilitate the reader first steps with vmf90.
The following programs are required: git~\cite{git-scm.com}, a Fortran compiler
[e.g. gfortran~\cite{gfortran-web}] and the HDF5 library.
\begin{verbatim}
# obtain vmf90 from its main repository
git clone https://github.com/pdebuyl/vmf90.git
# enter vmf90 directory
cd vmf90
# create the directory in which vmf90 is built
mkdir build
# enter the directory in which vmf90 is built
cd build
# build vmf90 for the hmf model
make -f ../scripts/Makefile hmf
\end{verbatim}
At this point, the program ``vmf90\_hmf'' is created within the ``build''
directory. Running the program is done by typing {\tt ./vmf90\_hmf} and requires
to have a configuration file ``HMF\_in'' in the directory.
Section~\ref{sec:sim-hmf} presents an example run for the HMF model.

\section{Algorithms}

\subsection{Semi-Lagrangian algorithm}

The semi-Lagrangian algorithm provides a way to solve numerically an advection
equation for data stored on a mesh. Instead of using finite difference
equations, the solution is found by following the trajectory of the transported
quantity backward in time. The semi-Lagrangian methodology does not impose a CFL
condition on the time step~\cite{sonnendrucker_et_al_semi-lag_1999}. The
semi-Lagrangian algorithm for the Vlasov equation is described in
Ref.~\cite{sonnendrucker_et_al_semi-lag_1999}. We reproduce here the steps of
the algorithm for clarity.

$f^s(x_i,v_m)$ is the distribution function at time step $s$ and grid point
$i,m$. $i$ is the spatial index and $m$ the velocity index. The time
discretization reads
\begin{equation}
  \label{eq:CandK}
  f^{s+1}(x,v) = f^s( x- \Delta t (v + \frac{1}{2} F^\ast(\bar x)\Delta t) , v -
  \Delta t F^\ast(\bar x)) ~,
\end{equation}
where $\bar x = x - v\Delta t /2$ and $F^\ast$ is the force field computed at half a time step.

A practical implementation to compute Eq. (\ref{eq:CandK}) on a numerical mesh is the following :
\begin{enumerate}
\item Advection in the $\theta$-direction, 1/2 time step\\
  $f^\ast(\theta_i,p_m) = f^s(\theta_i-p_m\Delta t/2, p_m)$.
\item Computation of the force field for $f^\ast$.
\item Advection in the $p$-direction, 1 time step\\
  $f^{\ast\ast}(\theta_i,p_m) = f^\ast(\theta_i, p_m - F^\ast(\theta_i) \Delta t)$.
\item Advection in the $\theta$-direction, 1/2 time step\\
  $f^{s+1}(\theta_i,p_m) = f^{\ast\ast}(\theta_i-p_m\Delta t/2, p_m)$.
\end{enumerate}

The evaluation of $f$ is done with cubic spline interpolation.

\subsection{Boundaries of the domain}

The interpolation is made to return the following quantities when a point
outside of the computational box is queried for interpolation:
\begin{itemize}
\item In a non-periodic box, $f$ is evaluated with zero value whenever
  \begin{equation}
    v \geq v_\textrm{max} \textrm{ or }
    v \leq v_\textrm{min} \textrm{ or }
    x \geq x_\textrm{max} \textrm{ or }
    x \leq x_\textrm{min}
  \end{equation}
\item In a periodic system, $x$ values outside of the box are evaluated at their
  periodic image inside the box. $v$ values outside the box are still evaluated
  to zero.
\end{itemize}
The evaluation as zero outside of the box corresponds to a zero-value
boundary condition. The box needs to be taken large enough so that the
normalization (conservation of the $L_1$ norm) is satisfied.

\section{Simulation of the Hamiltonian Mean-Field model}
\label{sec:sim-hmf}

For reference, we reproduce the first Vlasov simulations for the HMF model
\cite{antoniazzi_califano_prl} with a waterbag initial condition $M_0=0.5$,
$U=0.69$. The configuration file is found in the ``scripts'' directory of the
vmf90 software as ``HMF\_in.resonances''.

Assuming that you are in the directory ``build'', as in Sec.~\ref{sec:build},
the instructions are
\begin{verbatim}
# copy example configuration file
cp ../scripts/HMF_in.resonances ./HMF_in
# run simulation
./vmf90_hmf
\end{verbatim}
The program creates the file ``hmf.h5'' that contains the trajectory of the
system. One may use any HDF5-capable software to extract the information. We
illustrate here the use of the program ``show\_vmf90.py'' that is found in the
``scripts'' directory of vmf90.
\begin{verbatim}
# plots Mx and My as a function time
../scripts/show_vmf90.py hmf.h5 plot Mx My
# displays phase space snapshots
../scripts/show_vmf90.py hmf.h5 snaps
\end{verbatim}
The two above examples display information that can be saved in a graphics file
format (as supported by matplotlib, e.g. eps, pdf, png and more). If one wishes
to dump the data for analysis with a tool that lacks HDF5 support,
``show\_vmf90.py'' features a ``dump'' command.
\begin{verbatim}
../scripts/show_vmf90.py hmf.h5 dump Mx My
\end{verbatim}
dumps $t$, $M_x$ and $M_y$ as text into the terminal. This text can be piped to
another program or dumped into a file by appending {\tt >~t\_Mx\_My.txt} to
the command line above.

\begin{figure*}
  \centering
  \begin{minipage}{.49\linewidth}
    \includegraphics[width=\linewidth]{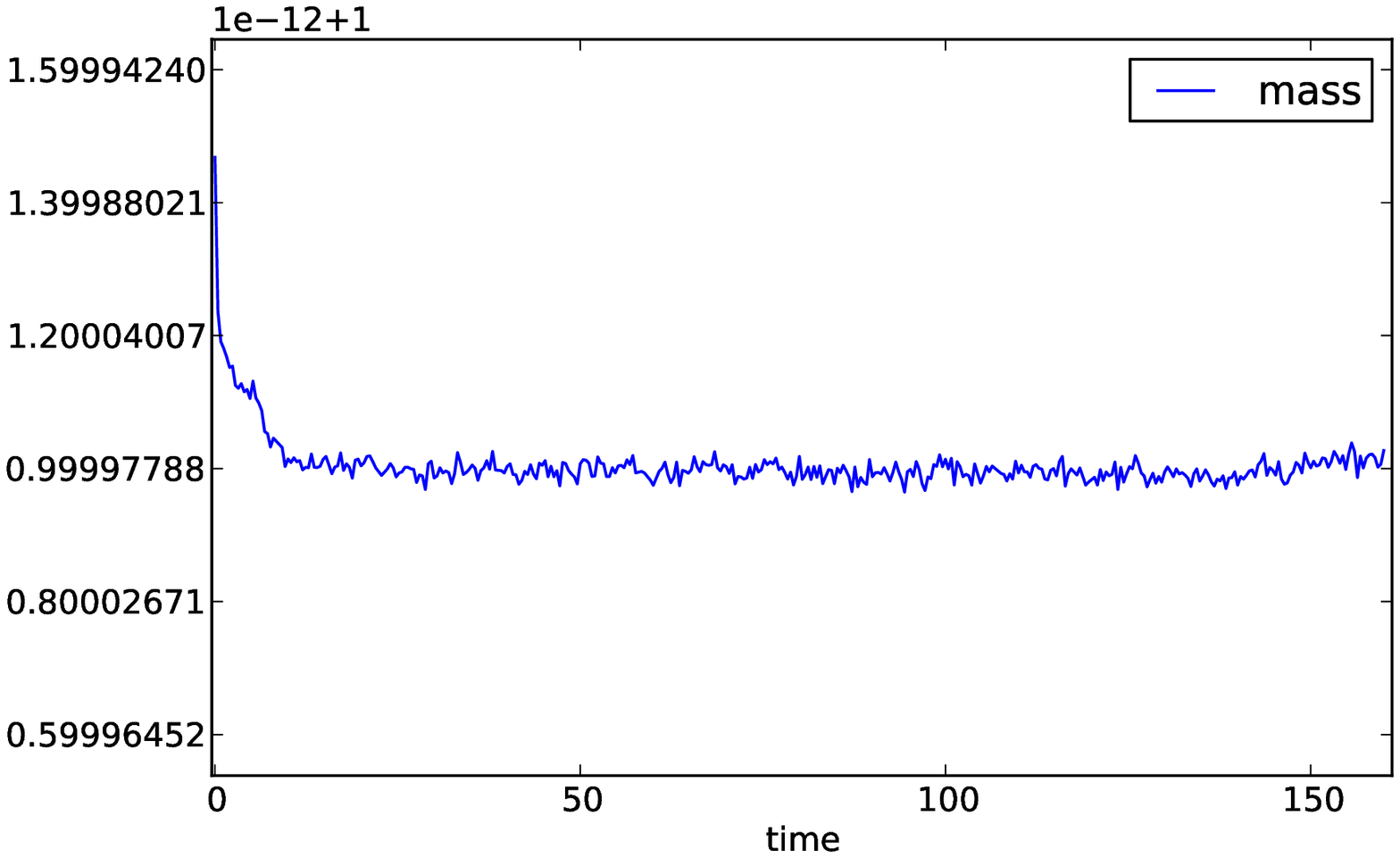}
  \end{minipage}
  \begin{minipage}{.49\linewidth}
    \includegraphics[width=\linewidth]{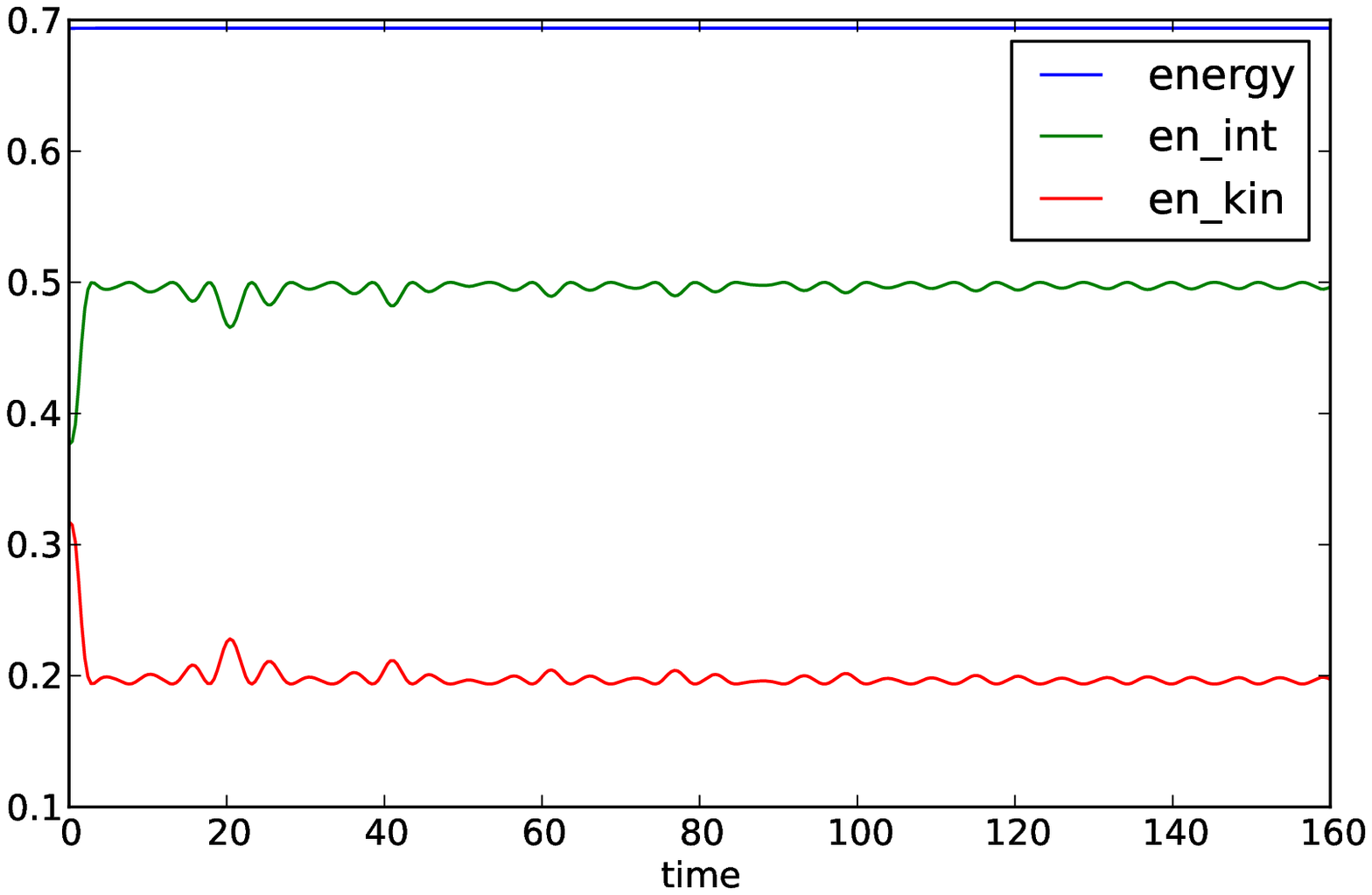}
  \end{minipage}

  \begin{minipage}{.49\linewidth}
    \includegraphics[width=\linewidth]{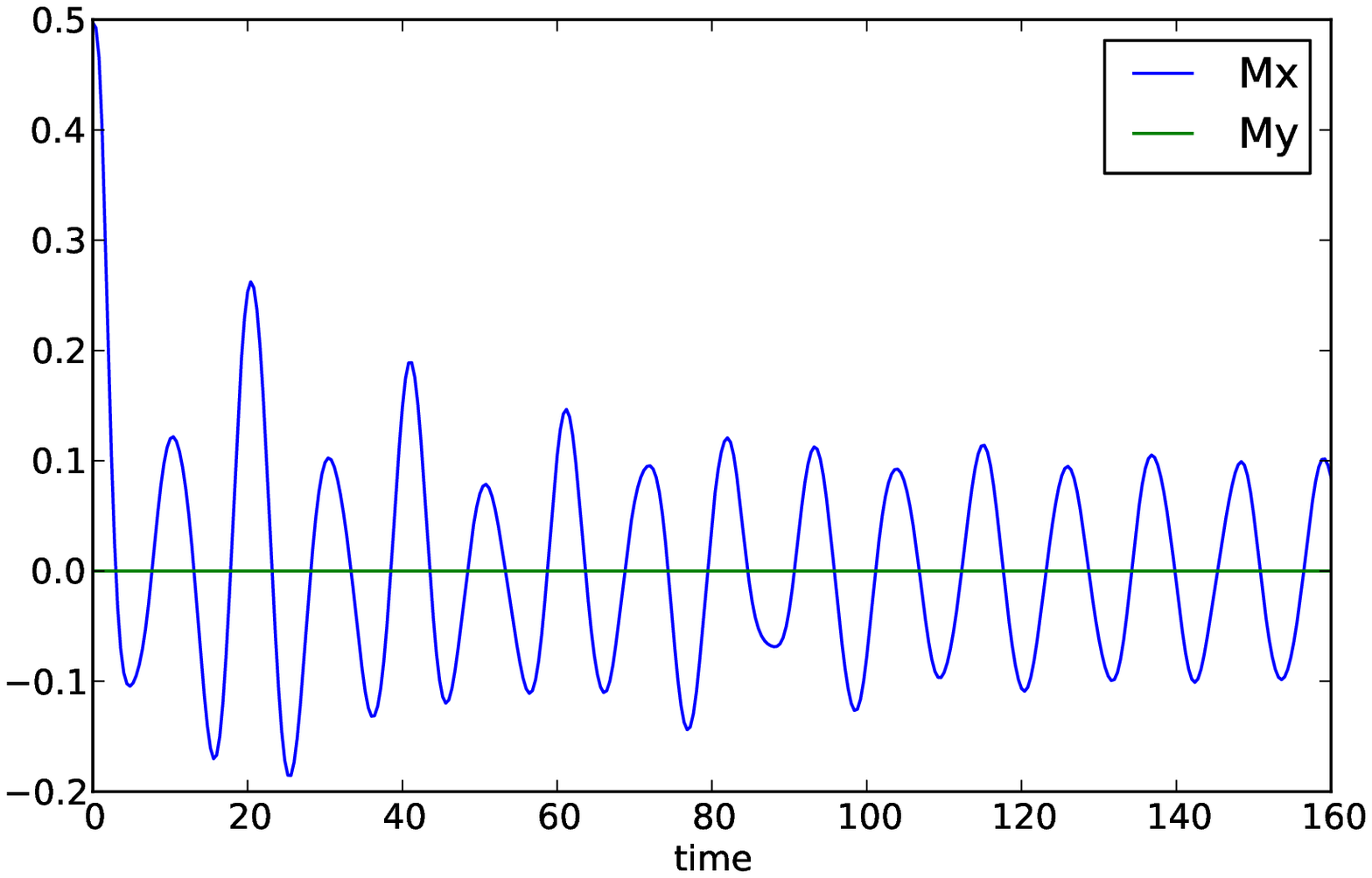}
  \end{minipage}
  \begin{minipage}{.49\linewidth}
    \includegraphics[width=\linewidth]{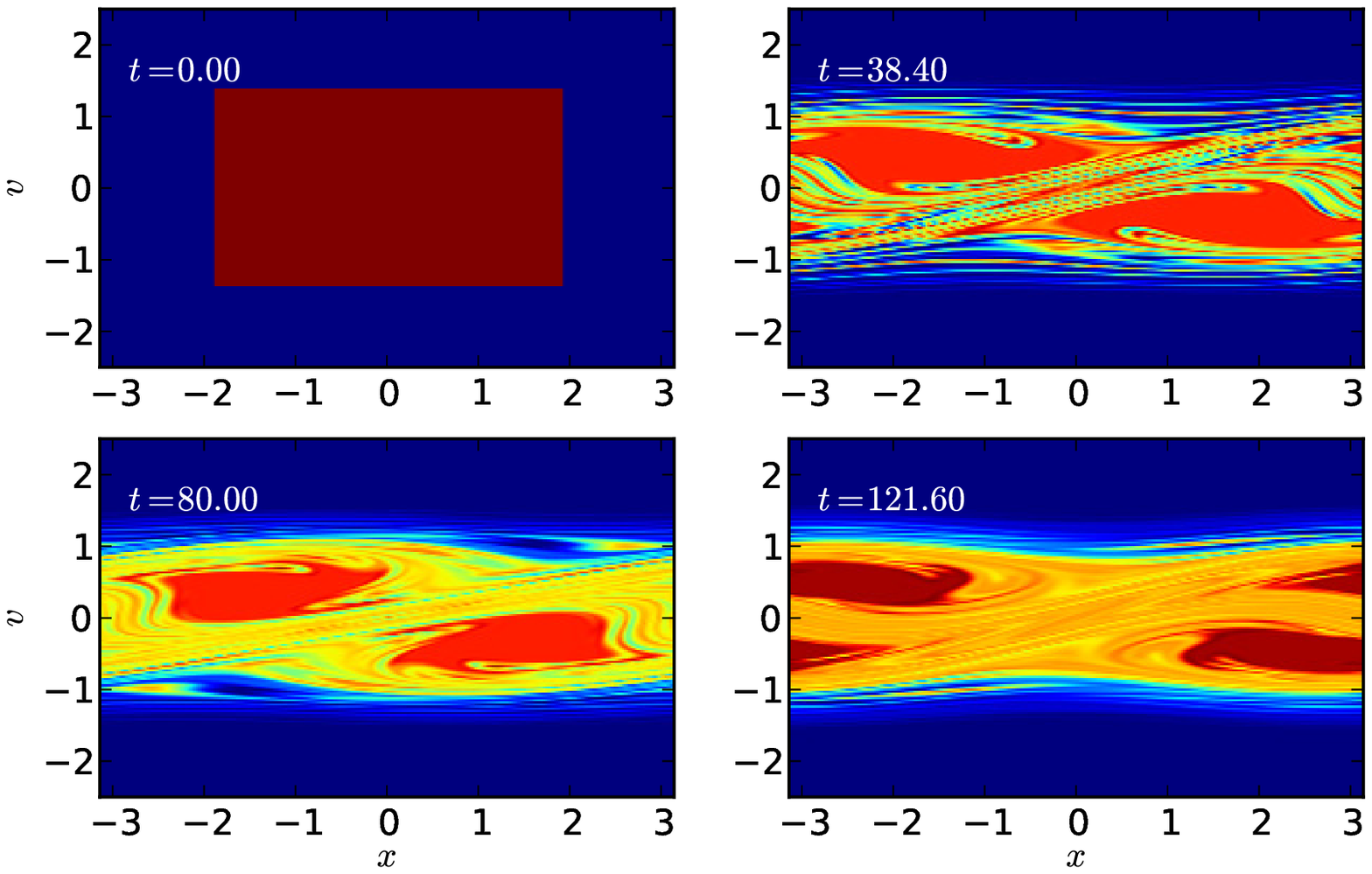}
  \end{minipage}

  \caption{Simulations of the ``HMF\_in.resonances'' example file. The mass
    evolution is zoomed to $1$ with a detail of $10^{-12}$. These figures are
    produced, with no modification, by the script ``show\_vmf90.py'' that
    accompanies vmf90.
    \label{fig:HMF_res}}
\end{figure*}
The figures generated by ``show\_vmf90.py'' are reproduced in
Fig.~\ref{fig:HMF_res} without further processing. One can already observe the
major features of the simulation: the normalization (``mass'', here) is
respected to about $10^{-12}$ and the energy is conserved. The behavior of the
magnetization, starting from $M_x=M_0=0.5$, is to decrease and then display
sustained oscillations. Snapshots of the phase space are also displayed, at four
regularly spaced times of the simulation, and the clustering behavior observed
in Refs.~\cite{antoni_ruffo_1995,antoniazzi_califano_prl} is well reproduced.

\subsection{Performing a parametric study}

vmf90 comes with an example program, ``parametric\_run.py'', in the
``scripts'' directory to perform a parametric study of the HMF magnetization.
Such a study represents a typical use of the program in the existing literature.

The example reproduces Fig.~3 of Ref.~\cite{de_buyl_et_al_cejp_2012}, albeit
with a smaller number of grid points and a shorter sampling time, for a faster
execution. The appearance of the figure is also changed to a scatter plot to
make the implementation of the data collection more readable.

\section{Simulations for the single-wave model}

\begin{figure*}
  \centering
  \begin{minipage}{.49\linewidth}
    \includegraphics[width=\linewidth]{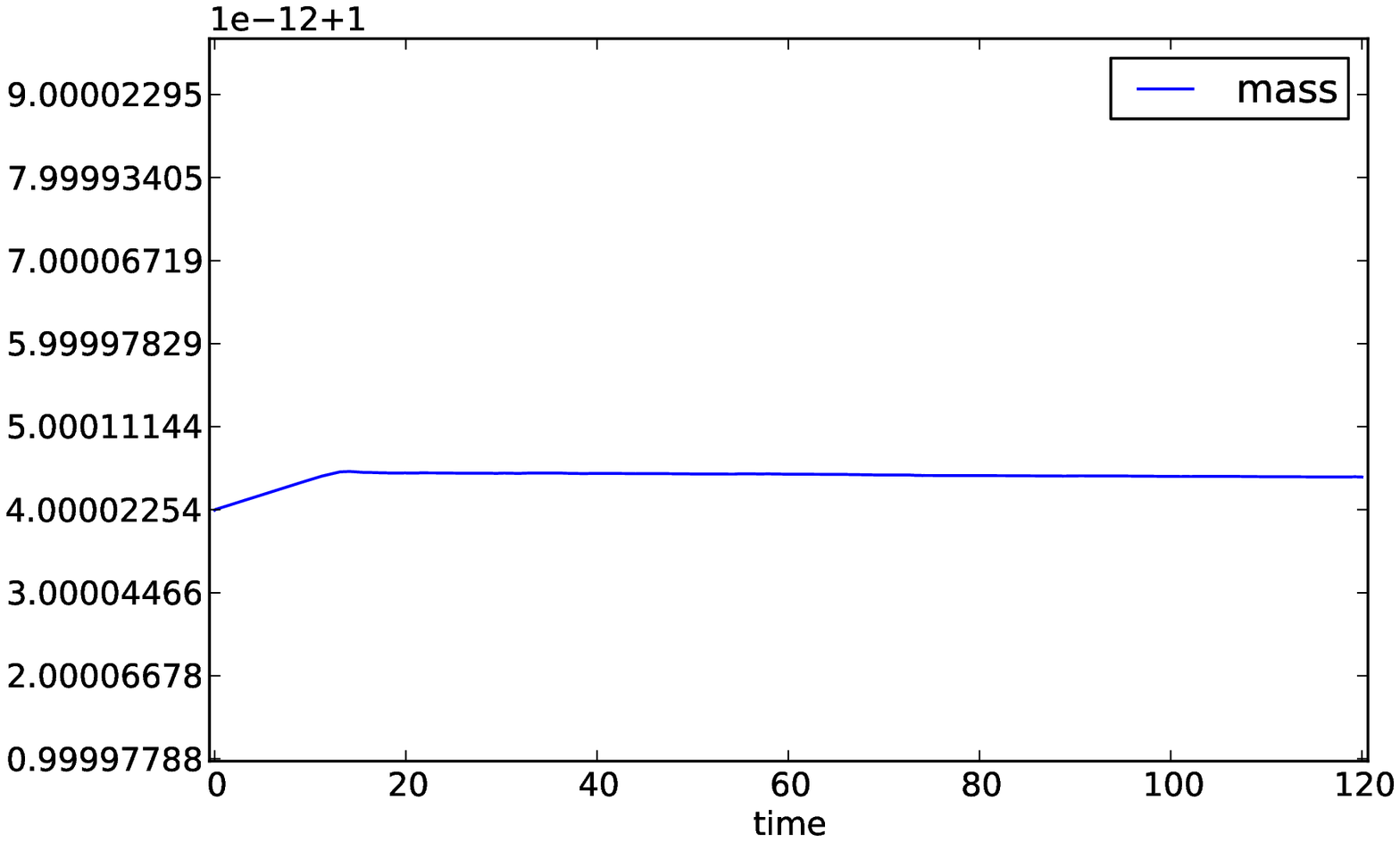}
  \end{minipage}
  \begin{minipage}{.49\linewidth}
    \includegraphics[width=\linewidth]{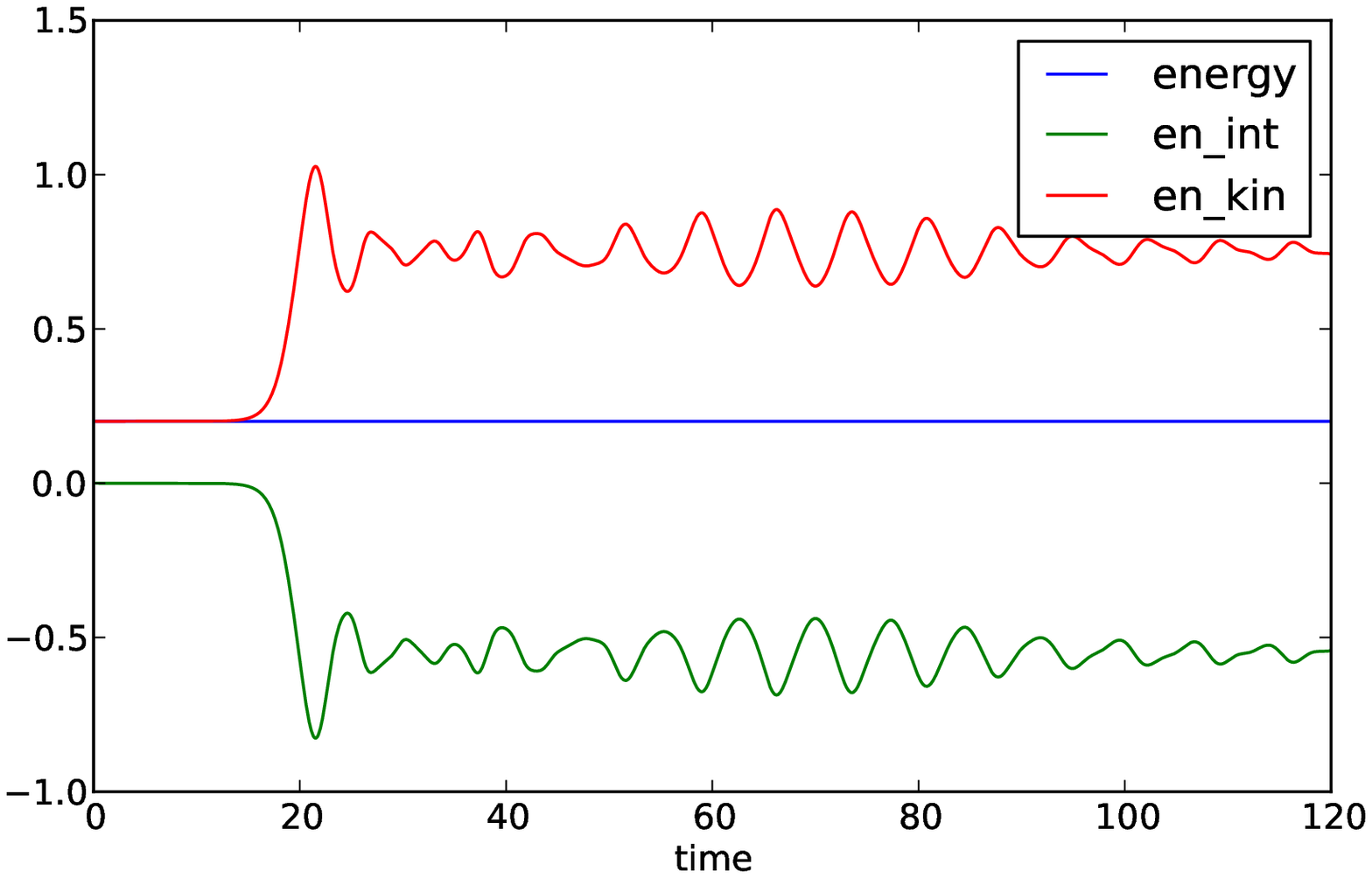}
  \end{minipage}

  \begin{minipage}{.49\linewidth}
    \includegraphics[width=\linewidth]{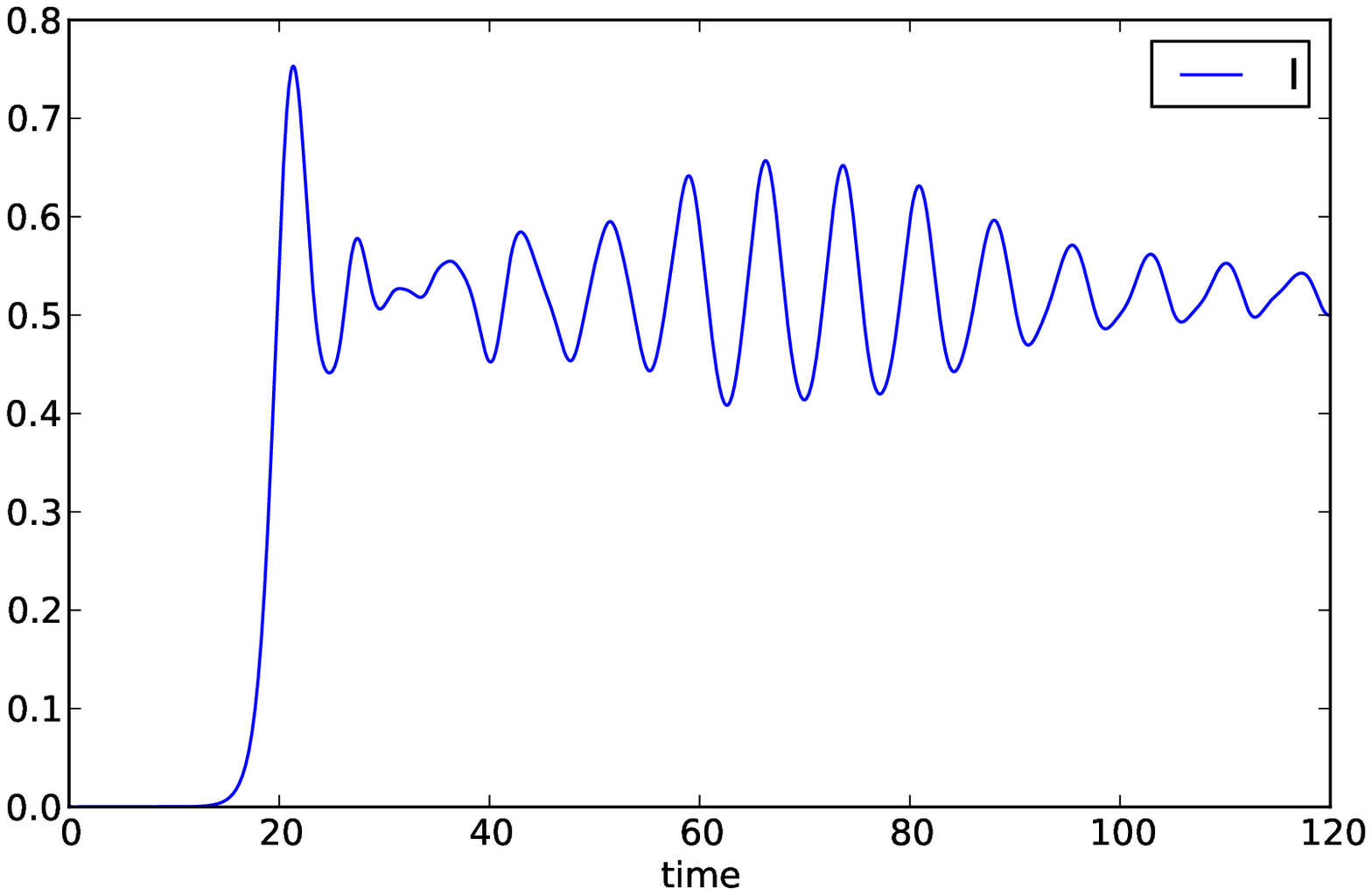}
  \end{minipage}
  \begin{minipage}{.49\linewidth}
    \includegraphics[width=\linewidth]{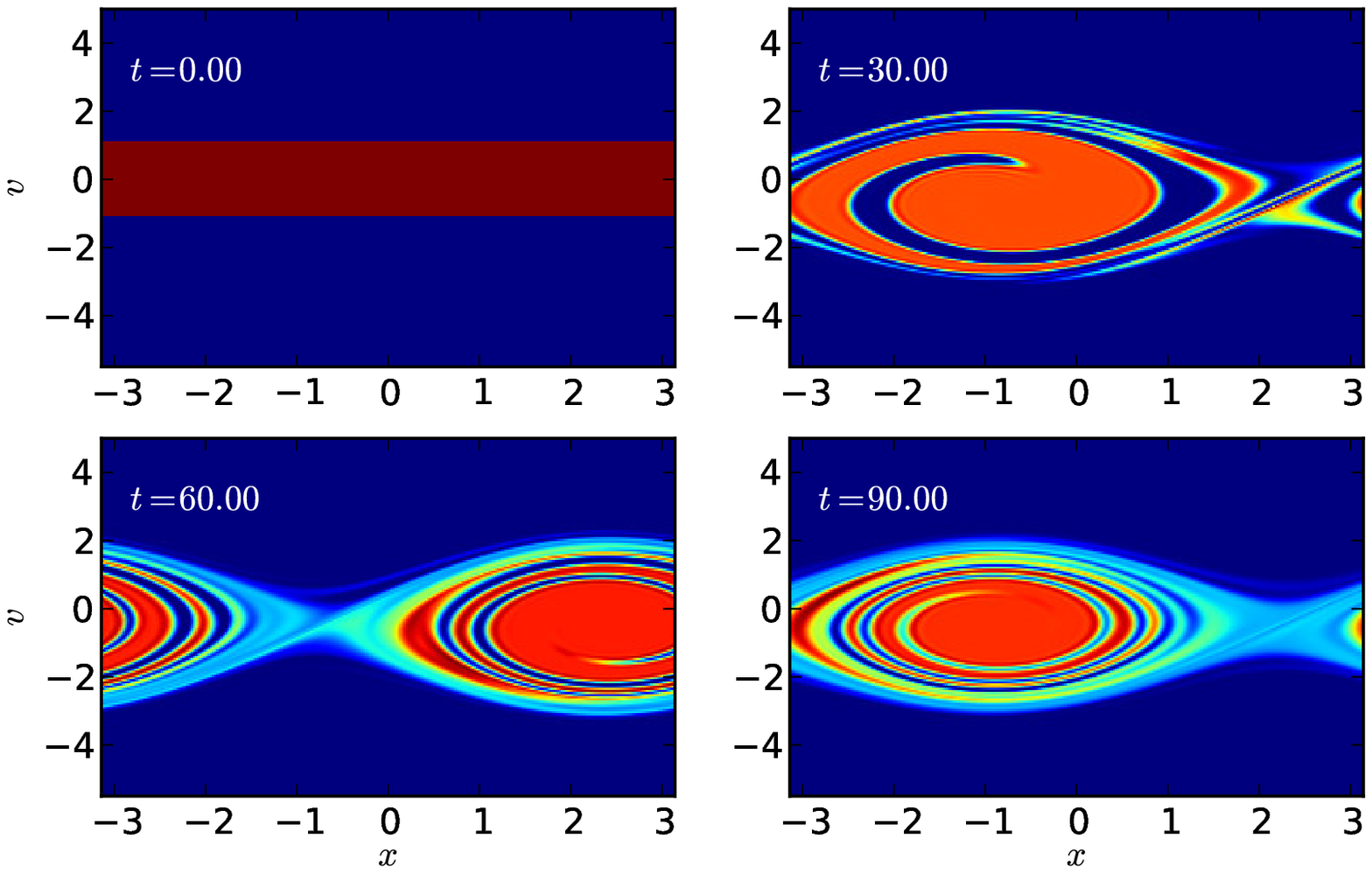}
  \end{minipage}

  \caption{Simulations of the ``FEL\_in'' example file. The mass
    evolution is zoomed to $1$ with a detail of $10^{-12}$. These figures are
    produced, with no modification, by the script ``show\_vmf90.py'' that
    accompanies vmf90.
    \label{fig:FEL_res}}
\end{figure*}

The single-wave model~\cite{OWM_pof_1971} represents an ensemble of particles in
which the interaction is well represented by a single component of the
electromagnetic field.
In addition to the particles, represented by the distribution function $f$, one
needs to consider also the field $A_x$, $A_y$ in the simulation as evidenced by
the coupled evolution equations~(\ref{eq:vlasovwave}).
As is typical, we track the evolution of the intensity of the field
$I=\sqrt{A_x^2+A_y^2}$.

We consider a simulation that was presented in
Ref.~\cite{de_buyl_et_al_prstab_2009} with a homogeneous water bag initial
condition of energy $U=0.2$ and a ``seed'' intensity given by $A_x=0.0001$,
$A_y=0$. Without the seed, even unstable initial conditions may remain stuck
because of the lack of granularity of the distribution.
The panels of Fig.~\ref{fig:FEL_res} are the direct output of the program
``show\_vmf90.py''.
Mass conservation is respected with a relative accuracy of about $10^{-12}$, as
it was for the HMF model.
The other panels display the evolution of the energy and of the intensity of the
wave. The snapshot displays the typical structure of a single ``bump'' in which
the majority of the distribution function is found.

\section{Discussion}

We have presented an open-source implementation of the semi-Lagrangian algorithm
for Vlasov simulations tailored for mean-field models. This code has served as
the basis for several publications and its modular nature allows for an easy
addition of other models.
The Fortran code uses modern additions to the language that allows for structured
programming.
Other technical features include the use of HDF5 for the storage of simulation
data, the inclusion of a simulation analysis program, revision control of the
source code and an extensive in code documentation.

\section*{Acknowledgments}

The author acknowledges fruitful collaboration with all of his co-authors listed
in the references.
The author also acknowledges Peter~H.~Colberg for many useful technical
discussions, Claire Revillet and Jakub Krajniak for testing the compilation.

\appendix

\section{Example configuration file}
\label{sec:config}

An example configuration file is provided, for the HMF model, in
Fig.~\ref{fig:config}. This file is available as ``HMF\_in.stable\_wb'' in the
program repository. Each parameter appears with an explicit denomination in the
file.

\begin{figure*}
\begin{verbatim}
model = HMF     ! the model name. is checked in the program 
Nx = 256        ! number of points in x-direction 
Nv = 512        ! number of points in v-direction
vmax = 2.5      ! size of the box
DT = 0.1        ! timestep
n_steps = 5     ! number of steps in inner loop
n_top = 300     ! number of executions for inner loop
n_images = 50   ! number of dumps of the phase space DF
IC = wb_eps     ! initial condition
width = 4.      ! \Delta\theta for initial condition
bag = 0.72      ! \Delta p for initial condition
epsilon = 0.001 ! initial perturbation of the profile
Nedf = 0        ! number of points for energy DF
\end{verbatim}
\caption{Example configuration file ``HMF\_in''.\label{fig:config}}
\end{figure*}

\bibliographystyle{elsarticle-num}
\providecommand{\urlprefix}{} 
\bibliography{../../biblio/2012_ucl,../../biblio/books,vmf90_2011}

\end{document}